\pdfoutput=1
\documentclass[a4paper,12pt]{article}
\usepackage{amsmath,amssymb,amsfonts}
\usepackage[multiple,stable]{footmisc}
\usepackage[amsmath,amsthm,thmmarks]{ntheorem}
\allowdisplaybreaks[4]
\usepackage[noconfig]{refstyle}
\usepackage[dvipsnames]{xcolor}
\usepackage[hyperfootnotes=false, linktocpage=true, colorlinks, citecolor=blue, linkcolor=blue, urlcolor=Maroon]{hyperref}
\usepackage{array,tabularx,booktabs,geometry,subfigure,graphicx,longtable}
\usepackage[bf]{caption}
\def\bZ{\mathbb{Z}}

\def\bP{\mathbb{P}}

\def\cA{\mathcal{A}}

\def\cD{\mathcal{D}}
\def\cL{\mathcal{L}}
\def\cM{\mathcal{M}}
\def\cN{\mathcal{N}}
\def\cO{\mathcal{O}}
\def\cP{\mathcal{P}}
\def\cV{\mathcal{V}}
\def\cT{\mathcal{T}}

\usepackage{authblk}

\title{Machine Learning on Generalized \\
Complete Intersection Calabi-Yau Manifolds}

\author[1,2]{Wei Cui \footnote{weicui@bimsa.cn}}
\author[3]{Xin Gao \footnote{xingao@scu.edu.cn}} 
\author[1,2]{Juntao Wang \footnote{ juntao.wang@bimsa.cn}}

\affil[1]{Yau Mathematical Sciences Center, Tsinghua University, Beijing 100084, China}
\affil[2]{Beijing Institute of Mathematical Sciences and Applications, Beijing 101408, China}
\affil[3]{College of Physics, Sichuan University, Chengdu 610065, China} 
\date{}

\begin{document}
\maketitle	

\begin{abstract}
\noindent

Generalized Complete Intersection Calabi-Yau Manifold (gCICY) is a new construction of Calabi-Yau manifolds established recently. However, the generation of new gCICYs using standard algebraic method is very laborious. Due to this complexity, the number of gCICYs and their classification still remain unknown. In this paper, we try to make some progress in this direction using neural network. The results showed that our trained models can have a high precision on the existing type $(1,1)$ and type $(2,1)$ gCICYs in the literature. Moreover, They can achieve a $97\%$ precision in predicting new gCICY which is generated differently from those used for training and testing. This shows that machine learning could be an effective method to classify and generate new gCICY.

\end{abstract}

\thispagestyle{empty}
\setcounter{page}{0}
\newpage

\tableofcontents

\section{Introduction}

The perturbative superstring theories are required to have an underlying ten dimensional spacetime, which must be compactified down to four dimensions in order to describe the real world. 
The methods for doing so are best understood for supersymmetric compactifications, where without turning on more general fluxes, the compactification manifold must be a Calabi-Yau threefold ($X$).  One of the main methods to construct the Calabi-Yau manifolds is realized as complete intersections of polynomial hypersurfaces  embedded  in a well known \lq\lq ambient space" $\cA$, i.e, products of projective spaces, abbreviated here as `CICY's. These CICY threefolds were first discussed and classified in a series of articles in the 1980's \cite{Hubsch:1986ny,Green:1986ck,Candelas:1987kf,Candelas:1987du}. The most  tractable descriptions of CICY database was obtained in \cite{Anderson:2017aux}, where all the divisor classes of the CY threefold are simply inherited from the ones of the ambient space and hence a favorable description.  

Recently a novel extension and generalization, called generalized complete intersections (gCI) has been introduced in constructing Calabi-Yau manifold \cite{Anderson:2015iia}, known as gCICY.  It has attracted lots of study from the perspective of string phenomenology and also pure mathematics since its invention. In gCICY,  some of those hypersurfaces have a negative degree over some factors in the ambient space $\cal A$. 
It is convenient to write the matrix in the form where columns containing only semi-positive integers in the left while columns containing negative integers in the right. For example, a typical gCICY configuration matrix is 
\begin{equation}\label{ConfMatgCICY}
X = \def\arraystretch{1.2}
\left[
\begin{array}{c||ccc|ccc} 
\bP^{n_1}&  a^{1}_{1} & \cdots &a^{1}_{K_A} & b^1_1 & \cdots & b^1_{K_B} \\
\bP^{n_2} &  a^{2}_{1} & \cdots &a^{2}_{K_A} & b^2_1 & \cdots & b^2_{K_B}\\ 
\vdots &  \vdots &\ddots&\vdots & \vdots & \ddots & \vdots \\
\bP^{n_m} & a^{m}_{1} & \cdots&a^{m}_{K_A}  & b^m_1 & \cdots & b^m_{K_B} \\ 
\end{array}
\right]  \ , 
\end{equation}
where the entries in the columns $\{a_{\alpha}\}$, $\alpha=1,2,\ldots, K_A$ contain only semi-positive integers and the entries in the columns $\{b_{\beta}\}$ $\beta=1,2,\ldots, K_B$ contains negative integers. As shown in \cite{Anderson:2015iia}, when certain conditions are satisfied, such configurations can also lead to well-defined CY manifolds and it was called $(K_A, K_B)$ type of gCICY. In \cite{Anderson:2015iia}, $(1,1)$ and $(2,1)$ type of gCICY were constructed with some limitation. We will summarize such construction in section \ref{sec:gCICY}.
The diffeomorphism class and cohomology of such generalized complete intersections Calabi-Yau (gCICY) have been studied in \cite{Anderson:2015iia, Anderson:2015yzz, Berglund:2016yqo, Jia:2018iza} and was shown can be extended to non-Fano varieties \cite{Berglund:2016nvh, Berglund:2022dgb}. A rigorous scheme-theoretic definition of such generalized complete intersection is given in \cite{Garbagnati:2017rtb}.

However, as we will briefly show in  section \ref{sec:gCICY} that the construction of gCICY in general is very hard, depends nontrivially on the underlying manifold data and it presents an interesting challenge for machine learning \cite{Nielsen:2015nn, Ruehle:2020jrk, He:2022cpz}.
 Machine learning has been a good implement in theoretical physics research and leads to fruitful results during the last couple of years. With the help of machine learning people are able to deal with problems with more computational efficiency, especially the problems involving big data, for example, study the landscape of string flux vacua  \cite{Cole:2018emh,Cole:2019enn,Krippendorf:2021uxu,Cole:2021nnt, He:2021nag, Ruehle:2017mzq,Halverson:2019tkf, He:2020mgx, Bena:2021wyr, He:2021eiu}  as well as F-theory compactifications \cite{Carifio:2017bov,Wang:2018rkk,Bies:2020gvf}. This technique allows people to learn lots of quantities of Calabi-Yau manifolds, from its  toric building blocks like the orientifold structure   \cite{Gao:2021xbs} and triangulations \cite{Altman:2018zlc,Demirtas:2020dbm}, to the calculation of Hodge numbers \cite{Bull:2018uow,He:2018jtw,He:2020lbz,Erbin:2020tks}, line bundle 
coohomology \cite{Klaewer:2018sfl,Brodie:2019dfx}  and numerical metrics   \cite{Anderson:2020hux,Jejjala:2020wcc,Douglas:2020hpv,Larfors:2021pbb}.   
 
The newly established generalized Complete Intersection Calabi-Yau (gCICY) database  \cite{Anderson:2015iia}  are the ideal data for machine learning in several aspects.  
First, the  explicit formulas  to determine a gCICY  are not known and calculations rely on complicated and  computationally intense algorithms.  Second, in the original construction \cite{Anderson:2015iia}, the primary time constraints arises from the calculation of line bundle cohomology. Taking negative entries in the configuration matrices as varying parameters, the number of matrices needed to be tested and computation time for cohomology will increase drastically with the increasing of those parameters. So to complete the initial scan in finite time, the authors left for future study of configuraitons whose calculation time was greater than five minutes or the negative entries in this configuration is greater than 4. Third, other than the negative entries in the configuration matrices, there are the other two parameters in the construction of gCICYs, the type $(m,n)$ of configuration matrices. There is even no mechanism to tell us if $m$ or $n$ will stop increasing somewhere. So theoretically, what we need to test, without further guidance, is a dataset of infinite number of matrices. So in principle, it is almost formidable to get intuitions on classifying gCICYs by large scale scanning using traditional calculation tools. And this is exactly where we can exploit the power of machine learning on solving problems with large dataset and its power on prediciton. 

In our work here, we will use neural network as our first trial in this direction. For the possibility of getting a better machine learning model, we enlarge the existing dataset of type $(2,1)$ by finishing the calculation for all the matrices with negative entries larger that $-4$, by relaxing the time constraints put in \cite{Anderson:2015iia}. Based on the full type $(1,1)$ and the partial type $(2,1)$ data(with negative entry larger than $-4$), we trained a neural network model which can achieve at least $90\%$ precision on both validation and testing dataset. As we pointed out before, we need to use machine learning to tell us, at least partially, what will happen if we step out of what we already know, to predict something new. On the other hand, we should also have a rough idea of how precise the prediction could be. To do this, we first generate a dataset of $10,000$ gCICYS and $10,000$ non-gCICYS with negative entries $-5$ or $-6$ and then use our trained model on it to test the model's predicting power. What impressive is that, even though this dataset is generated randomly and quite different with the data for model training, the prediction precision of this model is $97\%$. This means that, the trained neural network is not over-fitting, grasping features which not only belong to those matrices with negative entries larger than $-4$ but also beyond. Guided by this result, we believe that the trained model should have already got some features which is not sensitive to the negative entries and can predict if a matrix represents a gCICY or not in a rather precise way. So together with this paper, we will attach a Mathematica file with all the predictions for matrices with negative entries $-5$ or $-6$.

This paper is organized as follows. In section \ref{sec:gCICY}, we briefly summarize the algorithm of how to construct a generalized Calabi-Yau manifold and also the basic idea of how to do its classification. In section \ref{MLgCICY} we will show in some detail how we train a neural network model by using data in type (1,1) and type (2,1) and then do a prediction. One thing in this section we want to emphasize is that, we did a prediction in a rather precise way, which is not always guaranteed in using machine learning. 
Finally, we make a conclusion in section \ref{sec:con}.

\section{The construction of gCICYs}
\label{sec:gCICY}

In this section we briefly summarize the construction of gCICYs.
An ordinary CICY $X$ is defined as a complete intersection of the zero loci of $K$ multi-homogeneous polynomials in the ambient space $A = \bP^{n_1} \times \bP^{n_2} \times \ldots \times  \bP^{n_m}$. 
The degrees of these polynomials are collected in a $m\times K$ matrix called {\em configuration matrix}, which is always used to represent the CICY X and it can be schematically written as:
\begin{equation}\label{ConfMatCICY}
X = \def\arraystretch{1.2}
\left[
\begin{array}{c||ccc} 
    \bP^{n_1}&  q^{1}_{1} & \cdots &q^{1}_{K} \\
    \bP^{n_2} &  q^{2}_{1} & \cdots & q^{2}_{K}\\
    \vdots &  \vdots &\ddots&\vdots\\
    \bP^{n_m} & q^{r}_{1} & \cdots& q^{r}_{K}\\ 
\end{array}
\right]  \ , 
\end{equation}
where each entry ${q^i}_a\in\mathbb{Z}^{\geq 0}$ specifies the degree of homogeneity of the $a^{\rm th}$ defining polynomial in the 
$i^{\rm th}$ projective ambient space and it should be a non-negative integer. In addition, these ${q^i}_a$'s should satisfy the Calabi-Yau condition $n_i+1=\sum_{a=1}^K{q^i}_a\quad\mbox{for}\quad i=1,\ldots ,m$.
If $X$ is a three-fold, then there should be one more condition that $\sum_i^m n_i=K+3$. 
The configuration matrix provides an efficient way to describe CICYs\cite{Hubsch:1986ny,Green:1986ck,Candelas:1987kf,Candelas:1987du}. 
The favorable description of the CICY database was obtained in \cite{Anderson:2017aux}, where all the divisor classes of the CY threefold are simply inherited from the ones of the ambient space \footnote{There exits a stronger notion of \lq\lq K\"ahler favourability" where  K\"ahler cones on $X$ descend  from an ambient space \cite{Anderson:2017aux}.   In some cases, the \lq\lq favorable" geometry are not \lq\lq K\"ahler
favorable" since the K\"ahler cone of $X$ is actually larger than the positive orthant one. }. Only about half of the descriptions in the original database \cite{Candelas:1987kf}  have this property.

Given a generalized configuration matrix introduced in (\ref{ConfMatgCICY}), the corresponding CY manifold, if exist, can be constructed in the following two steps. 
First,  consider the first $K_A$ columns containing only non-negative entries. Each one of these columns gives a degree $\{a_{\alpha}\}$ homogeneous polynomial. 
The complete intersection of the vanishing loci of these polynomials define a codimensional $K_A$ hypersurface $\cM$ in $\cA$ denoted by, 
\begin{equation}\label{ConfMatgCICYM}
\cM = \def\arraystretch{1.2}
\left[
\begin{array}{c||ccc} 
\bP^{n_1}&  a^{1}_{1} & \cdots &a^{1}_{K_A} \\
\bP^{n_2} &  a^{2}_{1} & \cdots & a^{2}_{K_A}\\
\vdots &  \vdots &\ddots&\vdots\\
\bP^{n_m} & a^{r}_{1} & \cdots& a^{r}_{K_A}\\ 
\end{array}
\right]  \ . 
\end{equation}

Second, consider the last $K_B$ columns in (\ref{ConfMatgCICY}). Unlike in the situation of regular CICYs, generally one cannot get a well-defined manifold using these negative valued columns. However, as observed in \cite{Anderson:2015iia}, manifolds can also be constructed using the rational functions instead of the homogeneous polynomials.
To be specific, we can define ${\cal L}_{\beta}=\cO_{\cM_{\beta-1}}(b^1_{\beta}\cdots, b^m_{\beta})$ to be the line bundle associated to the column $\{b_{\beta}\}$ with $\beta=1,2,\ldots, K_B$. 
Here $\{\cM_{\beta}\}$ is a sequence of nested hypersurface inside $\cM$ given by 
\begin{equation*}
    \cM=\cM_0 \xrightarrow{r_1=0} \cM_1 \xrightarrow{r_2=0} \cM_2 \xrightarrow{r_3=0} \ldots \xrightarrow{r_{K_B}=0} \cM_{K_B}=X
\end{equation*}
where $r_{\beta} \in H^0(\cM_{\beta -1 },L_{\beta})\neq 0$ are non-trivial global sections of each $L_{\beta}$ and are in general, not polynomials but rational functions. 
Thus, in order to have a well-defined manifold, 
we need to have nontrivial global sections, i.e. 
\begin{equation} \label{condgCICY1}
    h^0({\cal M}_{\beta-1},{\cal L}_{\beta})\neq 0, \quad \beta =1,2, \ldots, K_B .
\end{equation}
In this way, one can construct a gCICY $X$ from a general configuration matrix containing negative numbers. Notice that in the construction of gCICY described above, the order of $K_A$ semi-positive columns can be arbitrary while the order of $K_B$ columns is not random. A different order of these columns may lead to failure of the above condition.

The resulting gCICY $X$ in this construction is a codimension $K_A+K_B$ hypersurface in $\cA$. Thus, 
if we want $X$ to be an $N$-fold, one need to have $\sum n_i - K_A - K_B=3$. 
Since the columns of $K_A$ and columns $K_B$ play different roles in the construction, we will call the resulting manifold as type $(K_A,K_B)$ gCICY. 
In addition, to define a CY manifold, the entries in the configuration matrix needs to obey the Calabi-Yau condition i.e.  
\begin{equation} \label{gcicyCond}
    n_i+1 = \sum_{\alpha=1}^{K_A}{a^i}_{\alpha} + \sum_{\beta=1}^{K_B}{b^i}_{\beta}\quad\mbox{for} \quad i=1,2,\ldots, m.
\end{equation}
In this paper, we are mainly interested in the gCICYs three-fold. Without saying explicitly, all configuration matrix that will be considered in Sec.3 have already obeyed the three-fold condition and the Calabi-Yau condition.

By definition, every gCICY can be described by a configuration matrix described above, however, not every such matrix describes a gCICY. To be able to represent a gCICY, this matrix has to obey the cohomology condition expressed in (\ref{condgCICY1}), i.e. to check if a given matrix describe a gCICY, one needs to at least compute $K_B$ line bundle cohomology. Furthermore, to make sure the gCICY is well defined, one requires that the trivial line bundle cohomology on $X$ is
\begin{equation} \label{condgCICY2}
    h^*(X,\cO) = \{1,0,0,1\}.
\end{equation}
This condition is sufficient to avoid the non-reduced, non-connected and non-CY geometries \cite{Anderson:2015iia}. Thus, to check if a given configuration matrix describes a well-defined gCICY, one needs to compute at least $K_B+1$ line bundle cohomologies. However, the computation of line bundle cohomologies are in general laborious. Thus, our knowledge about gCICYs are still limited. In this paper, we will combine these limited knowledge and the powerful machine learning algorithm to train a binary classifier that can efficiently distinguish the configuration matrices that can describe gCICYs and those that cannot. We hope that our model can learn the essential knowledge about gCICY encoded in the configuration matrix and extend their power beyond the training data and find more gCICYs.



\section{Learning gCICYs and its prediction} \label{MLgCICY}

In this section, we will use the artificial neural network (ANN) to determine if a given configuration matrix defines a well-defined gCICY or not.   
We will do this for both type $(1,1)$ and type $(2,1)$ configuration matrices.

\subsection{Type $(1,1)$ gCICYs} \label{subsec:codim11}

The simplest gCICYs are given by configuration matrices with one column of positive entries and one column of negative entries like below

\begin{equation} \label{eq:X(1,1)}
X_{(1,1)}=
\left[\begin{array}{c||c|c}
\bP^{n_1} & a^{1} & b^{1}\\
\bP^{n_2} & a^{2} & b^{2}\\
\vdots & \vdots &\vdots\\
\bP^{n_m} & a^{m} & b^{m}\\
\end{array}\right] , 
\qquad
\mathcal{M}=
\left[\begin{array}{c||c}
\bP^{n_1} & a^{1} \\
\bP^{n_2} & a^{1} \\
\vdots & \vdots \\
n_m & a^{m} \\
\end{array}\right] \ .
\end{equation}
where the $a^i$'s are all semi-positive, whereas some of the $b^i$'s can take negative values. A configuration matrix $X$ defines a gCICY if it satisfies the two conditions described in Sec.2. 

The type $(1,1)$ gCICYs have been fully classified in \cite{Anderson:2015iia}. 
As discussed above, a configuration matrix $X$ describes a well-defined gCICY only if the cohomology $h^0(\mathcal{M},\mathcal{L})$ is non-trivial. 
This condition puts a bound on the entries in $X$. 
Using this bound,  one could find that there are totally $28$ type $(1,1)$ gCICYs 
in the ambient spaces
\begin{equation*}
    \bP^4\times \bP^1, \bP^3\times \bP^1\times\bP^1, \bP^2 \times \bP^2 \times \bP^1, \bP^2 \times \bP^1 \times \bP^1 \times \bP^1 , \bP^1 \times \bP^1 \times \bP^1 \times \bP^1 \times \bP^1
\end{equation*}
Among them, there are $16$ gCICYs that are given by definite configuration matrices while the configuration matrices for the rest $12$ depend on an arbitrary positive integer number $i \in \mathbb{Z}_{>0}$. For example, one of them has the form  
\begin{equation} \label{eqn:typeIIeg}
\left[\begin{array}{c||c|c}
\bP^{1} & 2+i& -i\\
\bP^{1} & 1 & 1\\
\bP^{3} & 0 & 4
\end{array}\right], \qquad i \in \mathbb{Z}_{>0} \ ,
\end{equation}
here each integer number $i$ will give us a configuration matrix, which may defines a gCICY. It seems that such configuration matrices lead to infinite number of gCICYS. 
However, as shown in \cite{Anderson:2015iia}, these infinite number of gCICYs can be either related to each other by the ineffective split transition (Type II/III) or they are just the multiple copies of the same manifold (Type I) 
\footnote{For the detailed discussion of different types of these equivalent classes, we refer the original paper \cite{Anderson:2015iia}}. 
Thus, there are only $28$ simply-connected type $(1,1)$ gCICYs.

In this paper, we will apply machine learning algorithms to study the classification of the gCICYs. 
Although type $(1,1)$ gCICYs have been fully classified, we will use them as our first example to show that machine learning is a viable and powerful way to solve such kind of mathematical problems. 

\subsubsection*{Data Generation}

As a first step for this task, we need to generate a proper dataset for the ANN. 
Since the maximum number of the rows in the product of projective spaces is $5$, the configuration matrices can be expressed as $5\times 2$ matrices. For matrices that are not in the size of $5\times 2$, we will pad them with zeros. 
Thus, our data will be in the form of $5\times 2$ with labels $1$ for gCICYs and $0$ for those that are not gCICYs.

The machine learning algorithms normally require big data. 
But we have only $28$ type $(1,1)$ gCICYs here. 
To construct a large enough data set, we can take more representatives in the 12 classes of indefinite configuration matrices including 
\begin{itemize}
    \item $7$ type II/III infinite classes.  
    For each class, we take three matrices, with $i=1,2,3$ respectively. 
    Although they correspond to the same gCICY, from the perspective of configuration matrices, they are different. 
    In this sense, they are useful for machine learning.

    \item $5$ type I infinite classes. 
    The different matrices in type I class will lead to multiple copies of the same gCICY. 
    The number of the copies is controlled by the entry $i$ in the matrices. 
    Since we are interested in the simply connected gCICYs, we will only take the matrix with $i=1$ in each class.  
\end{itemize}
Other than the above two kinds of matrices, we also put into our dataset the $16$ gCICYs represented by definite configuration matrix. So overall, we have $16+3\times 7+5 = 42$ different configuration matrices, which seems too less for a real machine learning. However, we can perform data enhancement by the permutation of rows and columns for these matrices and this will lead to a dataset with $2200$ matrices. 
All of them describe codimensional $(1,1)$ gCICYs and we will denote them by $M_{5\times 2}$.

Besides that, we also need another dataset of matrices $\Tilde{M}_{5\times 2}$ that do not correspond to gCICYs. 
These matrices should look similar with the configuration matrices of gCICYs so that one cannot distinguish them easily by simple principles without calculations. 
Besides the property of not being gCICY, we also require the $\Tilde{M}_{5\times 2}$ should satisfy the following conditions:
\begin{enumerate}
    \item $\Tilde{M}_{5\times 2}$ should obey the Calabi-Yau condition. 
    \item The first column contain only semi-positive integers and the second column contains at least one negative integer. 
    \item The entries of the first/second column in $\Tilde{M}_{5\times 2}$ should be in the range $[0,5]/[-3,5]$ respectively, which is the same range for entries in $M_{5 \times 2}$.  
    \item The matrices in $\tilde{M}_{5\times 2}$ do not correspond to gCICYs.
\end{enumerate}
A large set of configuration matrices satisfying all the above constraints can be constructed. 
We randomly choose $2200$ of them for machine learning and denote them by $\Tilde{M}_{5\times 2}$ in the following.

In summary, we have constructed $M_{5\times 2}$ consisting of $2200$ matrices describing gCICYs and also $\Tilde{M}_{5\times 2}$ consisting of $2200$ matrices that do not define gCICYs but otherwise are similar with those in $M_{5\times 2}$.
We will label them together as 
\begin{equation} \label{eqn:data_train_11}
    \cD =\{M_{5\times 2} \to 1\} \cup
 \{\Tilde{M}_{5\times 2} \to 0 \}\;, 
\end{equation}
%
and this is the dataset that we will use in the machine learning.

As is customary, we need to disjointly split the above dataset into a training set, a validation set and a test set. We typically use $10\%$ of the dataset for validation, and the remainder of it for training and testing. All of them are randomly generated from the full dataset.
The validation set is used to monitor learning progress during training process and the test set is used to evaluate the trained neural network. Now let us turn to more details on the machine learning part.

\subsubsection*{Learning type $(1,1)$ gCICY}

We will use a standard forward-feed, fully connected neural network for this task.
This neural network is a map from the input data, a $5\times 2$ matrix, to output data, a length-$2$ vector. 
The two components of the output vector summing to one can be understood as the probabilities for the matrix to define a gCICY or not.   
The architecture of the neural network is as follows 
\begin{equation} \label{eqn:NN_11}
    \cN_{10 \times 2} = (F \circ G_{10 \times 40},  F \circ G_{40 \times 20}, S \circ G_{20 \times 2})
\end{equation}
where $G_{n \times m}$ is a fully connected layer of neurons with $n$ inputs and $m$ outputs, $F$ is the standard Relu function and $S$ is a softmax layer transforming 
$2$ real numbers into a probability distribution of $2$ possible outcomes.

We implement this neural network using the PyTorch package\cite{pytorth:2019}. 
We use $90\%$ of the dataset ($3960$) for training and testing, $10\%$ of them ($440$) for validation. 
We will denote them by $\cT$ and $\cV$ respectively. 
The set $\cT$ is further divided into different portions for training and testing. More specifically, we will train our neural network with $10\%, 20\%, \ldots, 90\%$ of $\cT$ and test the trained model on the rest of $\cT$ and also the validation set $\cV$. 
\begin{figure}[!ht] 
\centering
\includegraphics[width=0.428\textwidth]{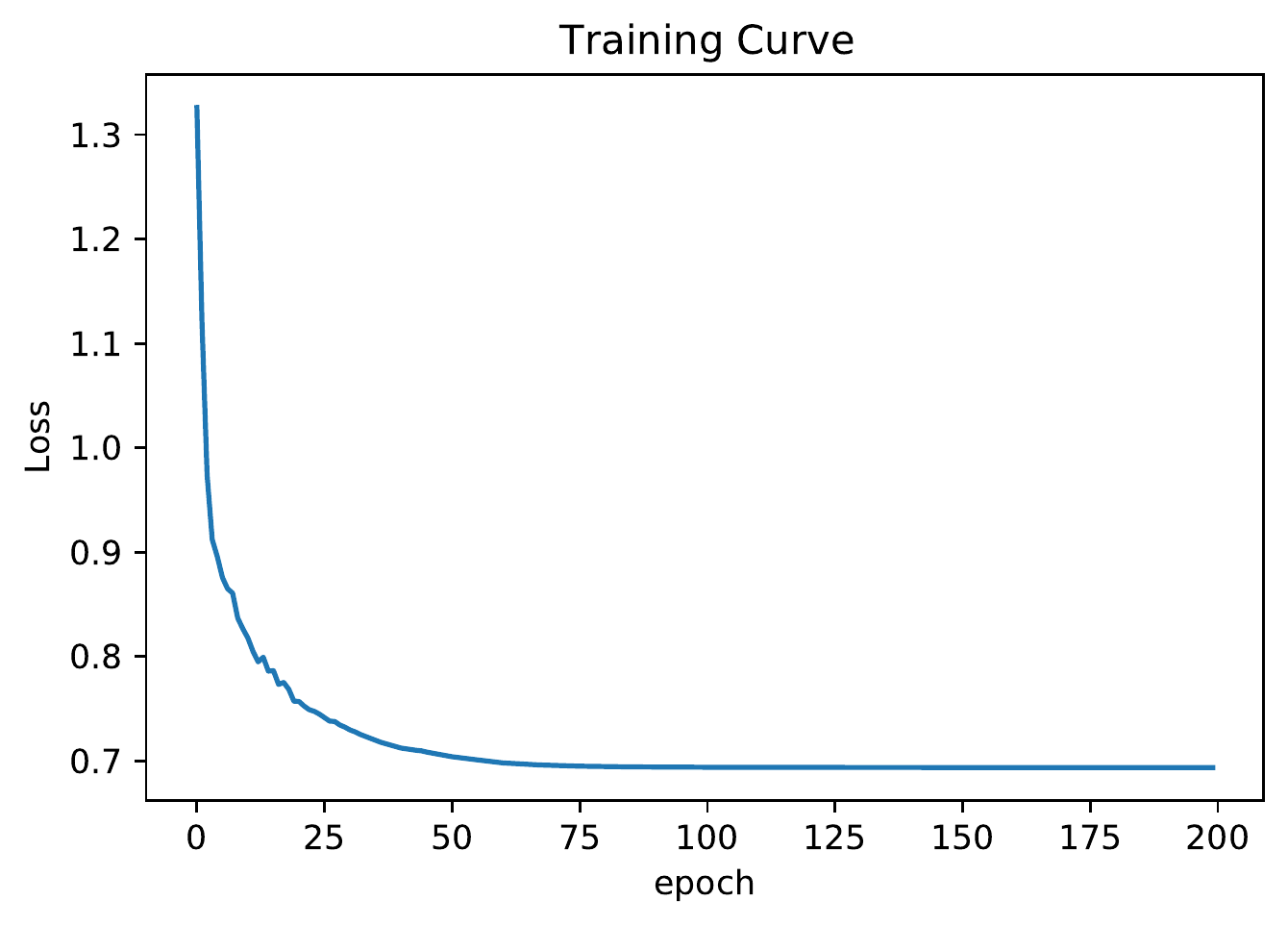}
\caption{The training curve when we take $10\%$ of $\cT$ for training.}
\label{fig:loss11}
\end{figure}
We will train the neural network defined in (\ref{eqn:NN_11}) with Adam optimizer and 'CrossEntropyLoss' loss function in $200$ epoches.
The initial learning rate is $0.025$ and we will reduce it by half after every $10$ epoches. 
In Fig.~\ref{fig:loss11}, we showed the learning result for using only $10\%$ of $\cT$. 
As we can see that the loss function converges nicely.  
Then, we evaluate our trained model and it can achieve $97.2\%$ accuracy on training set, $92.5\%$ on testing set and $93.4\%$ on validation set.

\begin{figure}[!ht] 
\centering
\subfigure{\includegraphics[width=0.428\textwidth]{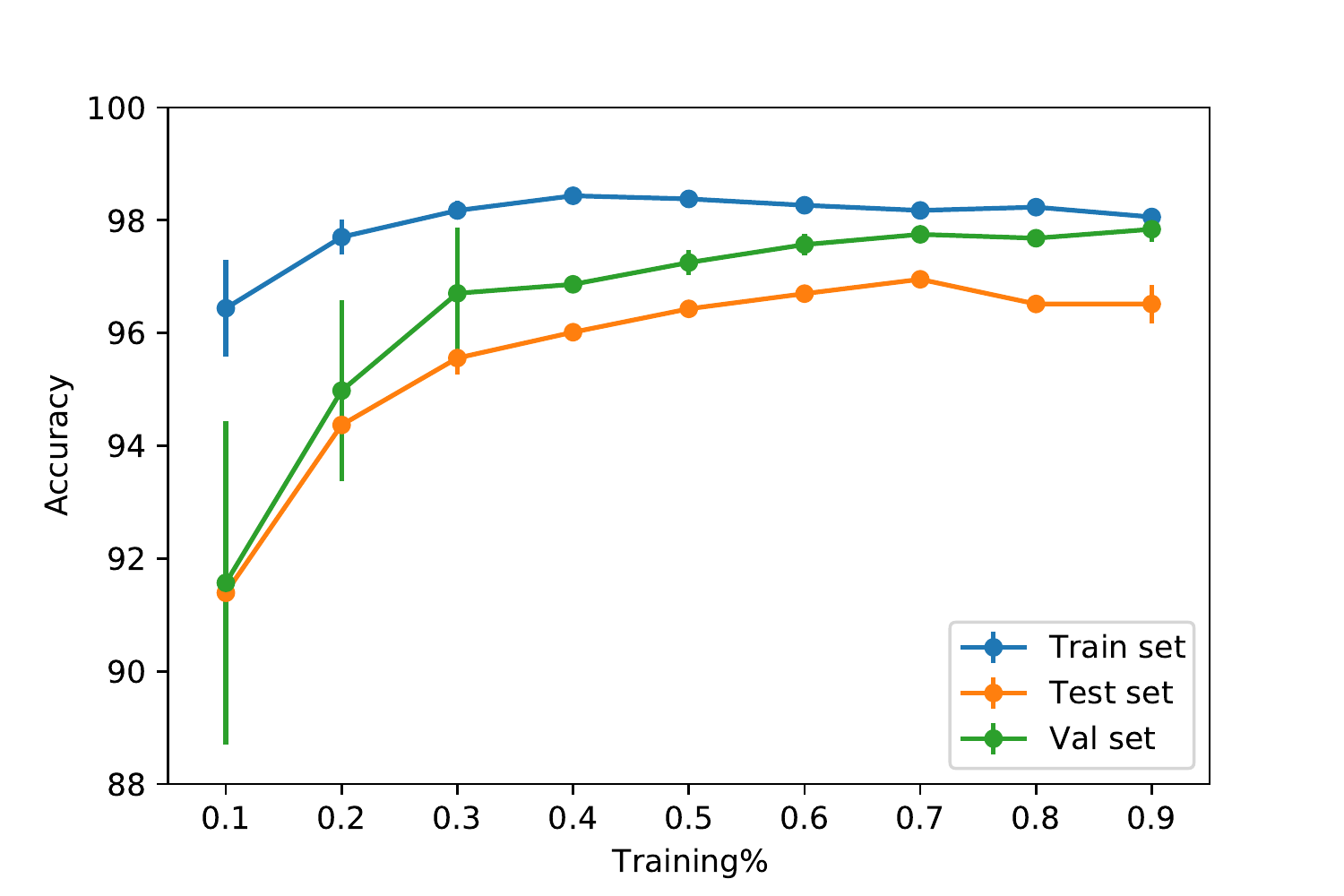}}
\subfigure{\includegraphics[width=0.428\textwidth]{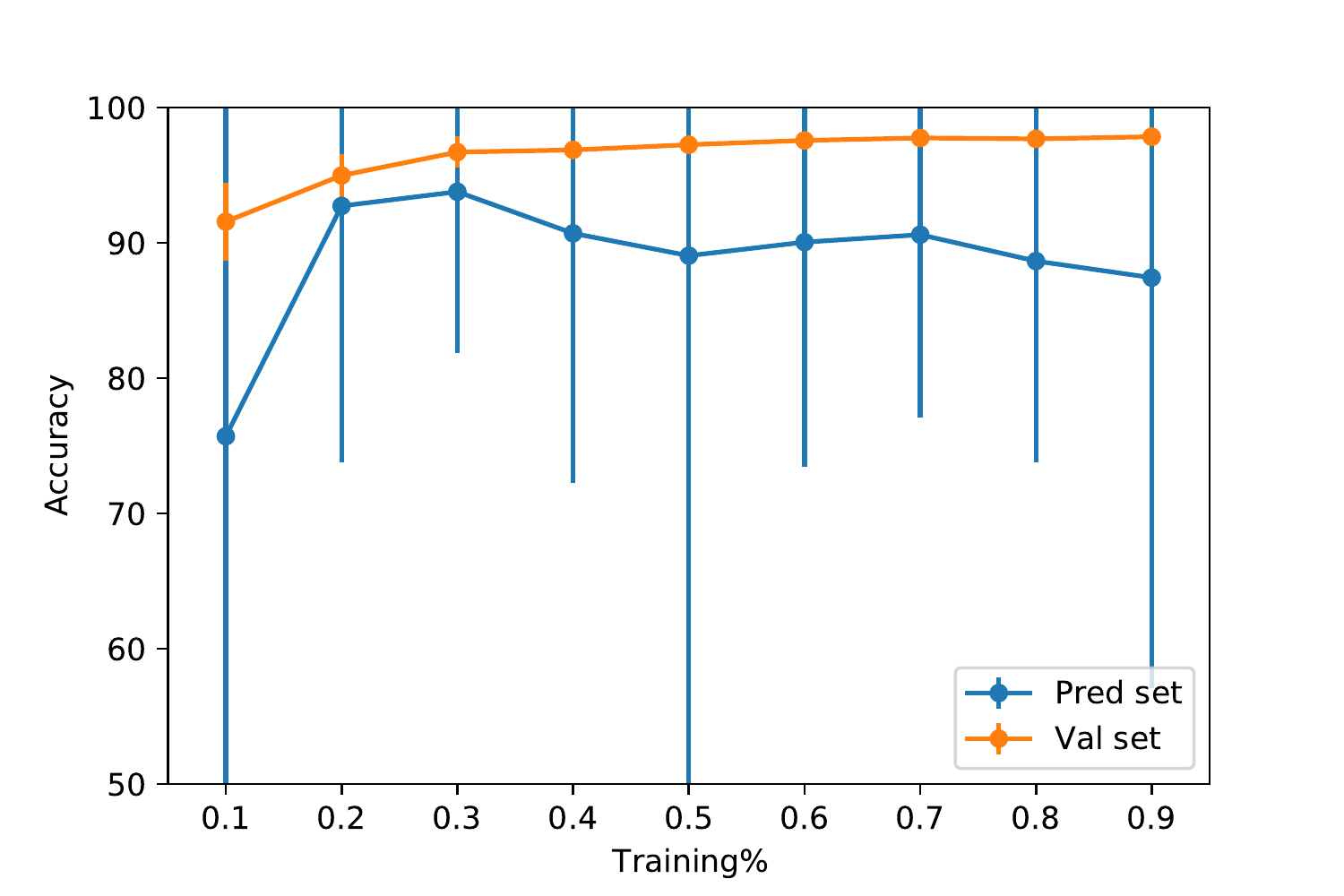}}
\caption{The accuracy of training/testing(left) and prediction(right) evaluated on models trained with different portion of $\cT$.
We keep the corresponding validation accuracy on both plots.}
\label{fig:acc/pred11}
\end{figure}

We also study the performance of our neural network trained with different portion of $\cT$. 
The complete result is presented in the left of Fig. \ref{fig:acc/pred11} where the error bar comes from the fact that each training set at a given percentage is randomly chosen 5 times and each of them give a different result. 
One can see that accuracy of the training, testing and validation increase with the size of training data and one can get as high as $96\%$ on testing set and $98\%$ on the validation set. 
This is a rather convincing performance by a relatively simple feed-forward network.

\subsubsection*{Prediction}

Besides the evaluation on the testing and validation set which are randomly chosen from the same dataset, it is more interesting to ask if our trained model can extend their classification power to a completely new dataset. For this purpose, we will construct more type $(1,1)$ gCICYs beyond the current dataset $\cD$ and test the predicting power of our trained model on this new dataset.

According to the classification of type $(1,1)$ gCICYs, one can generate more gCICYs by including more configuration matrices in the $7$ infinite classes of type II/III. 
We have considered the matrices with $i=1,2,3$ and also their permutations in $\cD$. 
To get more gCICYs, we can generate $7$ more matrices by simply taking $i=4$.
Even though these new matrices define the same gCICYs which are already covered by the matrices with $i=1,2,3$, we can still use them for prediction because it is not a prior that the trained machine learning model recognize them. 
After padding and data enhancement, we can obtain $400$ new $5 \times 2$ matrices and we denote this dataset by $\cP$, which is different from $\cD$ by construction.

%
%

Now, we can study the prediction power of our trained models using the data set $\cP$. 
The result is shown on the right of Fig. (\ref{fig:acc/pred11}). 
Again the error bar comes from the fact that the models are trained with $5$ different sets of data that are randomly chosen from $\cT$ in a fixed size.  
On average, we can see that our model can successfully predict at least $85\%$ of the gCICYs in the new dataset $\cP$. 
Therefore, we can say that the our machine learning model have learned the essential information to classify the gCICY and can be an useful tool beyond the training set.

\subsection{Type $(2,1)$ gCICYs}\label{21egs}


The type $(2,1)$ gCICYs are defined by the complete intersection of 2 hypersurfaces determined by $2$ polynomials and 1 hypersurface defined by a rational function. Their configuration matrices can generally be written as follows 
\begin{equation}\label{21general}
X_{(2,1)}=
\left[\begin{array}{c||cc|c}
\bP^{n_1} & a^{1}_1 & a^{1}_2  & b^{1}\\
\bP^{n_2} & a^{2}_1 & a^ {2}_2  & b^{2}\\
\vdots & \vdots &\vdots & \vdots\\
\bP^{n_N} & a^{m}_1 & a^{m}_2  & b^{m}\\
\end{array}\right] \ ,
\qquad
\mathcal{M}=
\left[\begin{array}{c||cc}
\bP^{n_1} & a^{1}_1 & a^{1}_2  \\
\bP^{n_2} & a^{2}_2 & a^ {2}_2 \\
\vdots & \vdots &\vdots \\
\bP^{n_N}& a^{m}_1 & a^{m}_2 \\
\end{array}\right] \ .
\end{equation}
where $a^{i}_1, a^{i}_2$ are semi-positive integers, $b^{i}$ are allowed to take negative integers and all the $a_i$ together with $b_i$ should satisfy the Calabi-Yau condition. For type $(2,1)$ gCICYs, there are only $10$ possible ambient spaces:
\begin{align*}
 &\bP^5 \times \bP^1, ~~ \mathbb{P}^4 \times \bP^2,~~\mathbb{P}^3 \times\bP^3,~~\mathbb{P}^4 \times\bP^1 \times\bP^1 \ , \\
 &\bP^3 \times\bP^2\times\bP^1,~~\mathbb{P}^2 \times\bP^2 \times\bP^2,~~\mathbb{P}^3 \times\bP^1 \times\bP^1 \times\bP^1 \ , \\
 &\mathbb{P}^2 \times\bP^2 \times\bP^1 \times\bP^1,~~\mathbb{P}^2 \times\bP^1 \times\bP^1 \times\bP^1 \times\bP^1,~~\mathbb{P}^1\times\bP^1 \times\bP^1 \times\bP^1 \times\bP^1 \times\bP^1 \ .
\end{align*}
As we will see later, there exists much more (might be infinite) type $(2,1)$ gCICYs compared with the type $(1,1)$ case.

As described in section \ref{sec:gCICY}, a gCICY configuration matrix should at least obey conditions (\ref{condgCICY1}) and (\ref{condgCICY2}), i.e. there should exist a non-trivial global section of $\cL$ $\equiv$ $ \cO_{\cM}(b^1,b^2, \ldots,b^m)$ and the trivial line bundle cohomology on $X$ should be $h^*(X,\cO)=\{1,0,0,1\}$. 
The configuration matrices satisfying both of the two conditions can be constructed as follows: 
\begin{itemize}
\item 
As shown in \cite{Anderson:2015iia}, to ensure the existence of global sections of $\cL$ on $\cM$, the negative components $b^i$ can only appear in the configuration matrix in the following patterns, in one $\mathbb{P}^1$ or two $\mathbb{P}^1$ factors, or in one $\mathbb{P}^2$.
%
%
Each matrix of this kind actually defines an infinite class of configuration matrices. 
We will call them 
{\it generalized configuration matrix}. 
In fact, one can show that for all the possible ambient spaces, there are totally $34,192$ generalized configuration matrices
\footnote{Notice that this is only a necessary condition. 
One still needs to compute $h^0(\cM,\cL)$ explicitly to make sure a given configuration matrix indeed defines a gCICY.}
. 

\item 
One also needs to impose the condition that $h^*(X,\cO)=\{1,0,0,1\}$. 
This can remove the Type I infinite class, but not the Type II/III 
\footnote{It is not clear how to remove the the Type II/III redundancies yet. 
There may still exist some redundancies in counting of these spaces like ineffective splits, some identity and reducedness, generally described in \cite{Anderson:2015iia}.}. 

\end{itemize}
Unfortunately, up to now there does not exist a classification for type $(2,1)$ gCICYs. 
However, we can still construct a subset of them with significant number of data which is sufficient for machine learning. 

Here, in order to give some sense to understand the first condition listed above(for a full proof please see \cite{Anderson:2015iia}) and also show the principle we use to generate more data for prediction, let us show the basic procedure to calculate line bundle cohomology and one of the techniques we use to select certain gCICY configuration matrices. 

The initial set up of our calculation is given by formula (\ref{21general}). We need to calculate the cohomology of $\mathcal{L}$, which is given by the last column of $X_{(2,1)}$:
\begin{equation}
\mathcal{L}=\mathcal{O}(b^{1},b^{2},\dots,b^{m}).
\end{equation}
More specifically, what we need to calculate is:
\begin{equation}
    h^0(\mathcal{M},\mathcal{L})
\end{equation}
and to do this we need to consider the following Koszul sequence:
\begin{equation}\label{koszul}
 0\longrightarrow \mathcal{N}_1^{*}\otimes\mathcal{N}_2^{*}\otimes\mathcal{L}\longrightarrow (\mathcal{N}_1^{*}\oplus\mathcal{N}_2^{*})\otimes\mathcal{L}\longrightarrow \mathcal{L}\longrightarrow \mathcal{L}|_{\mathcal{M}}\longrightarrow 0,   
\end{equation}
which could be further broken up into two short exact sequences:
\begin{equation}\label{koszul1}
 0\longrightarrow \mathcal{N}_1^{*}\otimes\mathcal{N}_2^{*}\otimes\mathcal{L}\longrightarrow (\mathcal{N}_1^{*}\oplus\mathcal{N}_2^{*})\otimes\mathcal{L}\longrightarrow \mathcal{K}\longrightarrow 0,   \end{equation}
 
\begin{equation}\label{koszul2}
 0\longrightarrow \mathcal{K}\longrightarrow \mathcal{L}\longrightarrow \mathcal{L}|_{\mathcal{M}}\longrightarrow 0,   
\end{equation}
where $\mathcal{K}$ is a carefully designed vector bundle to keep the exactness of the sequence. 
The long exact sequences of cohomology associated with the above two short exact sequences are given by 
\begin{eqnarray*}
&&0\longrightarrow h^0(\mathcal{A},\mathcal{N}_1^{*}\otimes\mathcal{N}_2^{*}\otimes\mathcal{L}) \longrightarrow h^0(\mathcal{A},(\mathcal{N}_1^{*}\oplus\mathcal{N}_2^{*})\otimes\mathcal{L})\longrightarrow h^0(\mathcal{A},\mathcal{K})\\
&&\longrightarrow  h^1(\mathcal{A},\mathcal{N}_1^{*}\otimes\mathcal{N}_2^{*}\otimes\mathcal{L})\longrightarrow h^1(\mathcal{A},(\mathcal{N}_1^{*}\oplus\mathcal{N}_2^{*})\otimes\mathcal{L})\longrightarrow h^1(\mathcal{A},\mathcal{K})\longrightarrow \dots. 
\end{eqnarray*}
and 
\begin{eqnarray*}
0\longrightarrow h^0(\mathcal{A},\mathcal{K}) \longrightarrow h^0(\mathcal{A},\mathcal{L})\longrightarrow h^0(\mathcal{M},\mathcal{L})
\longrightarrow  h^1(\mathcal{A},\mathcal{K})\longrightarrow h^1(\mathcal{A},\mathcal{L})\longrightarrow h^1(\mathcal{M},\mathcal{L})\longrightarrow \dots.
\end{eqnarray*}

To write the above complicated looking sequences in a concise way, we can put them into the following cohomology tables:
\begin{eqnarray}\label{cohtable1}
\begin{array}{cccc}
 \quad  & \mathcal{N}_1^{*}\otimes\mathcal{N}_2^{*}\otimes\mathcal{L} & (\mathcal{N}_1^*\oplus\mathcal{N}_2^*)\otimes\mathcal{L} & \mathcal{K}   \\
h^0 & a^0_1 & a^0_2 & a^0_3 \\
h^1 & a^1_1 & a^1_2 & a^1_3 \\
h^2 & a^2_1 & a^2_2 & a^2_3 \\
h^3 & a^3_1 & a^3_2 & a^3_3 \\
h^4 & a^4_1 & a^4_2 & a^4_3 
\end{array},
\end{eqnarray}
\begin{eqnarray}\label{cohtable2}
\begin{array}{cccc}
 \quad  & \mathcal{K} & \mathcal{L} & \mathcal{L}|_{\mathcal{M}}   \\
h^0 & a^0_3 & a^0_4 & a^0_5 \\
h^1 & a^1_3 & a^1_4 & a^1_5 \\
h^2 & a^2_3 & a^2_4 & a^2_5 \\
h^3 & a^3_3 & a^3_4 & a^3_5 \\
h^4 & a^4_3 & a^4_4 & a^4_5 
\end{array},
\end{eqnarray}
%
here each entry $a^i_j$ represents the cohomology of certain line bundle. For example, $a^0_1$ is just the zeroth cohomology of $\mathcal{N}_1^{*}\otimes\mathcal{N}_2^{*}\otimes\mathcal{L}$ on $\mathcal{A}$, $h^0(\mathcal{A},\mathcal{N}_1^{*}\otimes\mathcal{N}_2^{*}\otimes\mathcal{L})$. The basic idea of getting $h^0(\mathcal{M},\mathcal{L})$ is to first calculate cohomologies of certain line bundles on the product of projective spaces and then use maps from the above long exact sequences. From K\"unneth formula we can know that, on the product of projective spaces, a line bundle's cohomology can be written as the product form:
\begin{equation}
h^n(\mathbb{P}^{n_1}\times\dots\times\mathbb{P}^{n_m}, \mathcal{O}(q_1,\dots,q_m))=
\underset{k_1+\dots+k_m=n}{\bigoplus}
h^{k_1}(\mathbb{P}^{n_1},\mathcal{O}(q_1))\times\dots\times h^{k_m}(\mathbb{P}^{n_m},\mathcal{O}(q_m)).
\end{equation}

And on a single productive space, a line bundle's cohomology $h^q(\mathbb{P}^n,\mathcal{O}(k))$ is given by:
\begin{align}
h^q(\mathbb{P}^n,\mathcal{O}(k))  & = \begin{cases}
\displaystyle 
{k+n\choose n}~, & q=0, \quad k>-1 \\[8pt] 
\displaystyle 
1, & q=n, \quad k=-n-1\\[8pt]
\displaystyle
{-k-1\choose -k-n-1} ,~ & q=n, k<-n-1\\[8pt]
0, &\text{otherwise}
\end{cases}
\end{align}
One important thing here we will use heavily later is that, on a projective space $\mathbb{P}^n$, if the line bundle $\mathcal{O}$ has negative entry, then the only possible non-zero cohomology should be $h^n(\mathbb{P}^n,\mathcal{O})$.
Let us look at one example to see how the rewriting of cohomology table can help us to chase the calculation of cohomology and even let us to select certain examples in an algorithmic way. The configuration matrix we want to look at is:
\begin{equation}
    \left[\begin{array}{c||cc|c}
\bP^{1} & 0 & 1  & 1\\
\bP^{5} & 4 & 3  & -1\\
\end{array}\right]. \ 
\end{equation}
For this case, we need to calculate the cohomology of $\mathcal{O}(1,-1)$ on:
\begin{equation}
    M=\left[\begin{array}{c||cc}
\bP^{1} & 0 & 1  \\
\bP^{5} & 4 & 3  \\
\end{array}\right]. \ 
\end{equation}
The long exact sequence for this calculation is:
\begin{equation}
0\longrightarrow \mathcal{O}(0,-8)\longrightarrow\mathcal{O}(1,-5)\oplus \mathcal{O}(0,-3)\longrightarrow \mathcal{O}(-1,1)\longrightarrow \mathcal{O}(-1,1)|_M\longrightarrow 0,
\end{equation}
from which we can get the following cohomology table according to (\ref{cohtable1}):
\begin{eqnarray}\label{cohtable31}
\begin{array}{cccc}
 \quad  & \mathcal{O}(0,-8) & \mathcal{O}(1,-5)\oplus \mathcal{O}(0,-3) & K   \\
h^0 & 0 & 0 & a^0_5 \\
h^1 & 0 & 0 & a^1_5 \\
h^2 & 0 & 0 & a^2_5 \\
h^3 & 0 & 0 & a^3_5 \\
h^4 & 0 & 0 & a^4_5 \\
h^5 & 21 & 0 & a^5_5
\end{array}.
\end{eqnarray}
Here all the negative entries are on the $\mathbb{P}^5$ factor, which means that the only non-zero cohomologies should be in the $h^5$ row. By chasing the long exact sequence along with this table, it is not hard to see that the only non-zero cohomology of $K$ is $h^4(\mathbb{P}^1\times\mathbb{P}^5,K)=21$. Through a similar procedure, it is not hard to write down the other part of the cohomology table which corresponds to (\ref{cohtable2}):
\begin{eqnarray}\label{cohtable3}
\begin{array}{cccc}
 \quad  & K & \mathcal{O}(1,-1) & \mathcal{O}(1,-1)|_M   \\
h^0 & 0 & 0 & 0 \\
h^1 & 0 & 0 & 0 \\
h^2 & 0 & 0 & 0 \\
h^3 & 0 & 0 & 21 \\
h^4 & 21 & 0 & 0 \\
h^5 & 0 & 0 & 0
\end{array}.
\end{eqnarray}
From this example we can see that, $h^5$ of $\mathcal{O}(0,-8)$ finally maps to $h^3$ of $\mathcal{O}(1,-1)|_M$. More precisely, we have exactly two maps, $h^5(\mathbb{P}^1\times\mathbb{P}^5,\mathcal{O}(0,-8))$ first maps to $h^4(\mathbb{P}^1\times\mathbb{P}^5,\mathcal{O}(1,-5)\oplus \mathcal{O}(0,-3))$, and then to $h^3(M,\mathcal{O}(1,-1))$. So if we look at the position of the number 21 in the cohomology table from left to right, it seems like it is climbing a ladder, from the row of $h^5$ to the row of $h^3$. This phenomenon follows from the exactness of the exact sequence and the number of map, and it is actually determined by the cohomology table's column number(which is actually determined by the codimension of M). So from the analysis of this example, we can say that a configuration matrix in the following form,
\begin{equation}\label{15matrix}
    \left[\begin{array}{c||cc|c}
\bP^{1} & \dots & \dots  & \dots\\
\bP^{5} & \dots & \dots  & -n\\
\end{array}\right] \ 
\end{equation}
with $n>0$, won't give us a gCICY. The basic idea is, by using the Bott-Borel-Weil theorem and also the Kunneth formula, the non-zero cohomologies should all in row $h^5$ or row $h^6$. But given that there are as much two maps in the cohomology table, so we can at most have a non-zero $h^3$ or $h^4$ for $\mathcal{L}$ on $\mathcal{M}$. More explicitly, the first part of the cohomology table for (\ref{15matrix}) could be:
\begin{eqnarray}\label{cohtable4}
\begin{array}{cccc}
 \quad  & \mathcal{N}_1^{*}\otimes\mathcal{N}_2^{*}\otimes\mathcal{L} & (\mathcal{N}_1^*\oplus\mathcal{N}_2^*)\otimes\mathcal{L} & \mathcal{K}   \\
h^0 & 0 & 0 & 0 \\
h^1 & 0 & 0 & 0 \\
h^2 & 0 & 0 & 0 \\
h^3 & 0 & 0 & 0 \\
h^4 & 0 & 0 & z \\
h^5 & x & y & y-x+z
\end{array}, \quad z=\ker(x\rightarrow y)
\end{eqnarray}
or
\begin{eqnarray}\label{cohtable5}
\begin{array}{cccc}
 \quad  & \mathcal{N}_1^{*}\otimes\mathcal{N}_2^{*}\otimes\mathcal{L} & (\mathcal{N}_1^*\oplus\mathcal{N}_2^*)\otimes\mathcal{L} & \mathcal{K}   \\
h^0 & 0 & 0 & 0 \\
h^1 & 0 & 0 & 0 \\
h^2 & 0 & 0 & 0 \\
h^3 & 0 & 0 & 0 \\
h^4 & 0 & 0 & 0 \\
h^5 & 0 & x & x+u \\
h^6 & y & z & z-y+u
\end{array}, \quad u=\ker
(y\rightarrow z)
\end{eqnarray}
No matter in any case, the only possible non-zero cohomology for $\mathcal{K}$ could be at row $h^3$, $h^4$ or $h^5$. So without writing down the other part of the cohomology table and do the chasing, we can see clearly that there is no chance to get a non-zero $h^0(\mathcal{M},\mathcal{L})$, there is just no enough steps for any of the non-zero $h^i$ to "climb" up to the row of $h^0$ of $\mathcal{L}$ on $\mathcal{M}$. 

The above example just illustrates one of the tricks we used to rule out configuration matrices which won't give us gCICYs. In certain cases, there are also tricks which can tell us which matrices definitely give us gCICYs. Those tricks help us save a lot of calculation time in preparing data for machine learning. After writing down the cohomology table, it is very easy to implement these tricks into Mathematica and do the scanning in a quick way. But in order to not go too far away from the main line of this paper, we won't give more details on this discussion.

\begin{table}[!h]
\begin{center}
\begin{tabular}{|c|c|c|c|}\hline
Embedding & \# of classes of generalized & \# of spaces found  & \# of spaces found  \\
   projective spaces   & configuration matrices & in previous scan \cite{Anderson:2015iia} & in our scan  \\\hline\hline
     $\mathbb{P}^5\times\mathbb{P}^1$ & 168& 28 & 67 \\\hline   
     $\mathbb{P}^4\times\mathbb{P}^2$ & 210 & 6 & 9 \\\hline   
   $\mathbb{P}^4\times\mathbb{P}^1\times\mathbb{P}^1$ & 1,197 & 229 & 369 \\\hline   
   $\mathbb{P}^3\times\mathbb{P}^2\times\mathbb{P}^1$ &1,800 & 263 & 341 \\\hline
   $\mathbb{P}^2\times\mathbb{P}^2\times\mathbb{P}^2$ & 550 & 12 & 12 \\\hline
   $\mathbb{P}^3\times\mathbb{P}^1\times\mathbb{P}^1\times\mathbb{P}^1$ & 4,410 & 545 & 860 \\\hline
   $\mathbb{P}^2\times\mathbb{P}^2\times\mathbb{P}^1\times\mathbb{P}^1$ & 5,235 & 520 & 683 \\\hline
   $\mathbb{P}^2\times\mathbb{P}^1\times\mathbb{P}^1\times\mathbb{P}^1\times\mathbb{P}^1$ & 12,180 & 770 & 1098 \\\hline
   $\mathbb{P}^1\times\mathbb{P}^1\times\mathbb{P}^1\times\mathbb{P}^1\times\mathbb{P}^1\times\mathbb{P}^1$ & 8,442 & 360 & 523 \\\hline\hline
   Total & 34,192 & 2,733 & 3,962 \\\hline
\end{tabular}
\caption{The distribution of codimension $(2,1)$ gCICYs founded in products of projective spaces.}  \label{tab:(2,1)distri}
\end{center}
\end{table}

Following the previous work \cite{Anderson:2015iia}, we start with the $34,192$ generalized configuration matrices and allow the parameter to vary from $1$ to $4$.
In this way, one can obtain $65,148$ independent configuration matrices with the most negative entry greater or equal to $-4$.
Then, we need to compute the line bundle cohomology of $h^0(\cM,\cL)$ and $h^*(X,\cO)$ for these configuration matrices. 
Different from the previous scan in \cite{Anderson:2015iia}, we remove the computing time constraint and consider all these $65,148$ matrices. 
The computation of line bundle cohomology is in general time consuming and is formidable to do on a laptop, so we finish our scan on the 'DragonTooth' cluster at Virginia Tech. 
Our result yielded $3,962$ type $(2,1)$ gCICYs that include the $2,862$ gCICYs found in previous scan of \cite{Anderson:2015iia}.  The distribution of these gCICYs among different ambient spaces is given in Table.\ref{tab:(2,1)distri}. 
In summary, we construct the complete dataset for all codimension $(2,1)$ gCICYs with negative entries larger than $-5$.
The explicit calculation becomes more and more time consuming as we decrease the negative entries in the configuration matrices. 
In the later part of this section, we will try to use the trained model for matrices with negative entry greater than $-4$ to do some prediction on cases with negative entry smaller than $-5$, to see if it can give us some clue to resolve this classification problem in the future.

\subsubsection*{Data Generation}

We take the scan result described above to be the dataset for machine learning. 
In specific, we have $3,962$ configuration matrices as gCICYs and $61,116$ matrices as non-gCICYs.  
Notice that the latter $61,116$ matrices, except for the property of not being gCICYs, are in many ways similar to the former $3,962$ matrices. 
After all, both of them are generated from the same generalized configuration matrices with entries in the same range $[-4,7]$.
For this reason, those matrices form the ideal dataset for our classification problem.

To generate the dataset for machine learning, we will first pad all these matrices into the form of $6 \times 3$. 
This is the maximum dimension for matrices considered in our case. 
Furthermore, we can perform rows and columns (only first two) permutations to these matrices.  This leads to totally $2,046,858$ matrices for gCICYs and $42,660,174$ matrices for those which are not gCICYs. 
We randomly choose $100,000$ matrices from each of these two classes and label the gCICYs as $1$ and $0$ for those which are not gCICYs, and we denote this dataset as $\cD$.

\subsubsection*{Learning type $(2,1)$ gCICYs}

Similar with the type $(1,1)$ case, we will use forward-feed, fully connected neural network here. But, the structure of neural network is more complicated. It is given by 
\begin{equation}
    \cN_{18 \times 2} =(F \circ G_{18 \times 100},  F \circ G_{100 \times 60}, F \circ G_{60 \times 30}, F \circ G_{30 \times 15}, F \circ G_{15 \times 8}, S \circ G_{8 \times 2})
\end{equation}
where the input is a $6 \times 3$ matrix while the output is a length-$2$ vector representing the probabilities for the matrix to be gCICY or not gCICY. 
We train this network in $200$ epoches. 
The initial learning rate is $0.01$ and will be reduced by $70\%$ after every $10$ epoches. 
The optimizer is Adam and the loss function is 'CrossEntropyLoss'. 
 
From the dataset $\cD$, we randomly choose $90\%$ of them for training and test, denoted by $\cT$. The rest of $10\%$ of them ($20,000$) will be used for validation denoted by $\cV$. Again, we will train our neural network with $10\%, 20\%, \ldots, 90\%$ of $\cT$ and evaluate it on $\cV$ every epoch.

\begin{figure}[!ht] 
\centering
\includegraphics[width=0.428\textwidth]{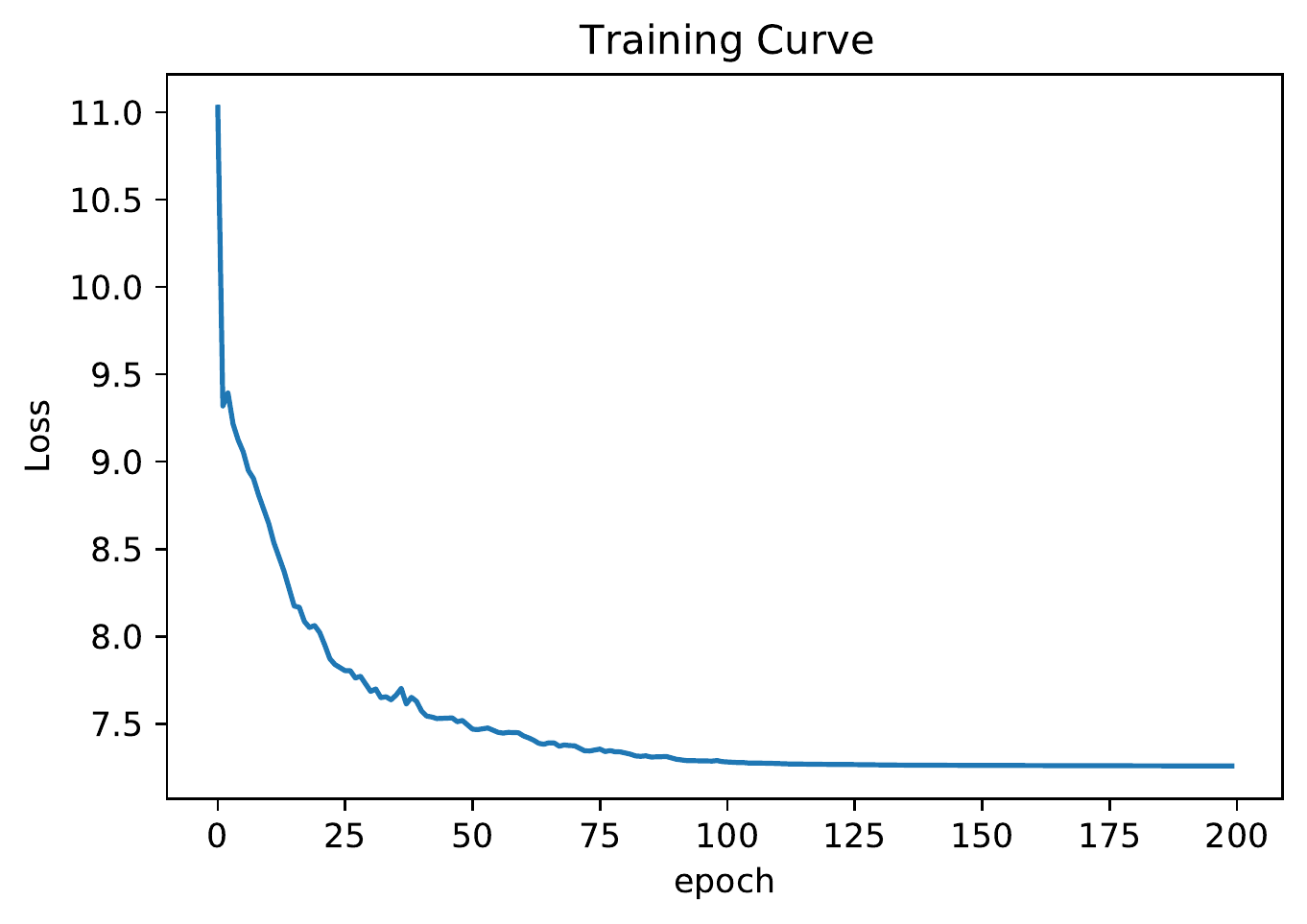}
\caption{The training curve when we take $60\%$ of $\cT$ for training.}
\label{fig:loss22}
\end{figure}

We implement this neural network with the Pytorch package and train it with the dataset generated above. 
In Fig.\ref{fig:loss22}, we plot the training curve for the model trained with $60\%$ of $\cT$.  As can be seen from this figure, the the loss function converges nicely. The trained model can achieve $95.1\%$, $92.3\%$ and $92.4\%$ on training, testing and validation set.

\begin{figure}[!ht] 
\centering
\subfigure{\includegraphics[width=0.428\textwidth]{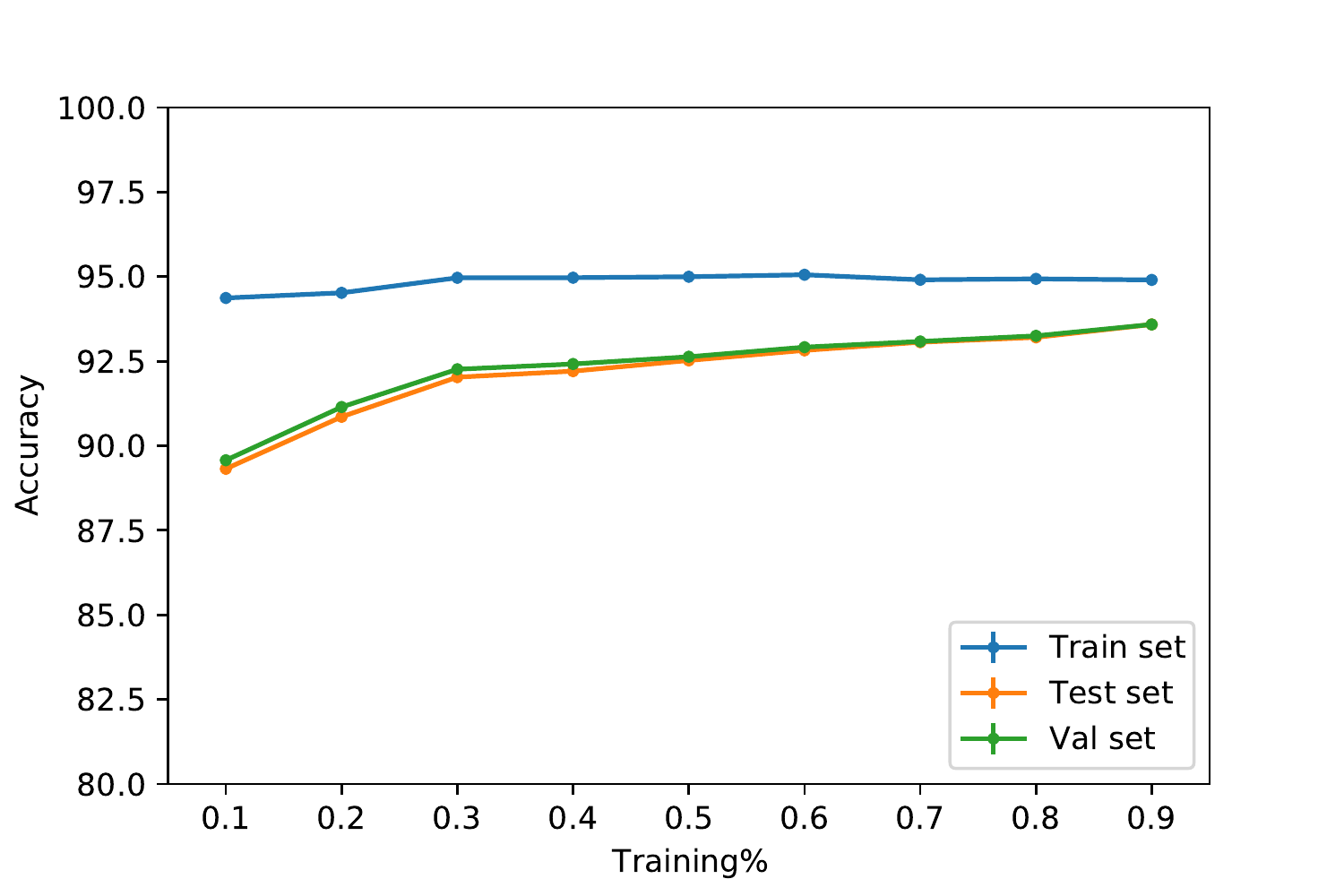}}
\subfigure{\includegraphics[width=0.428\textwidth]{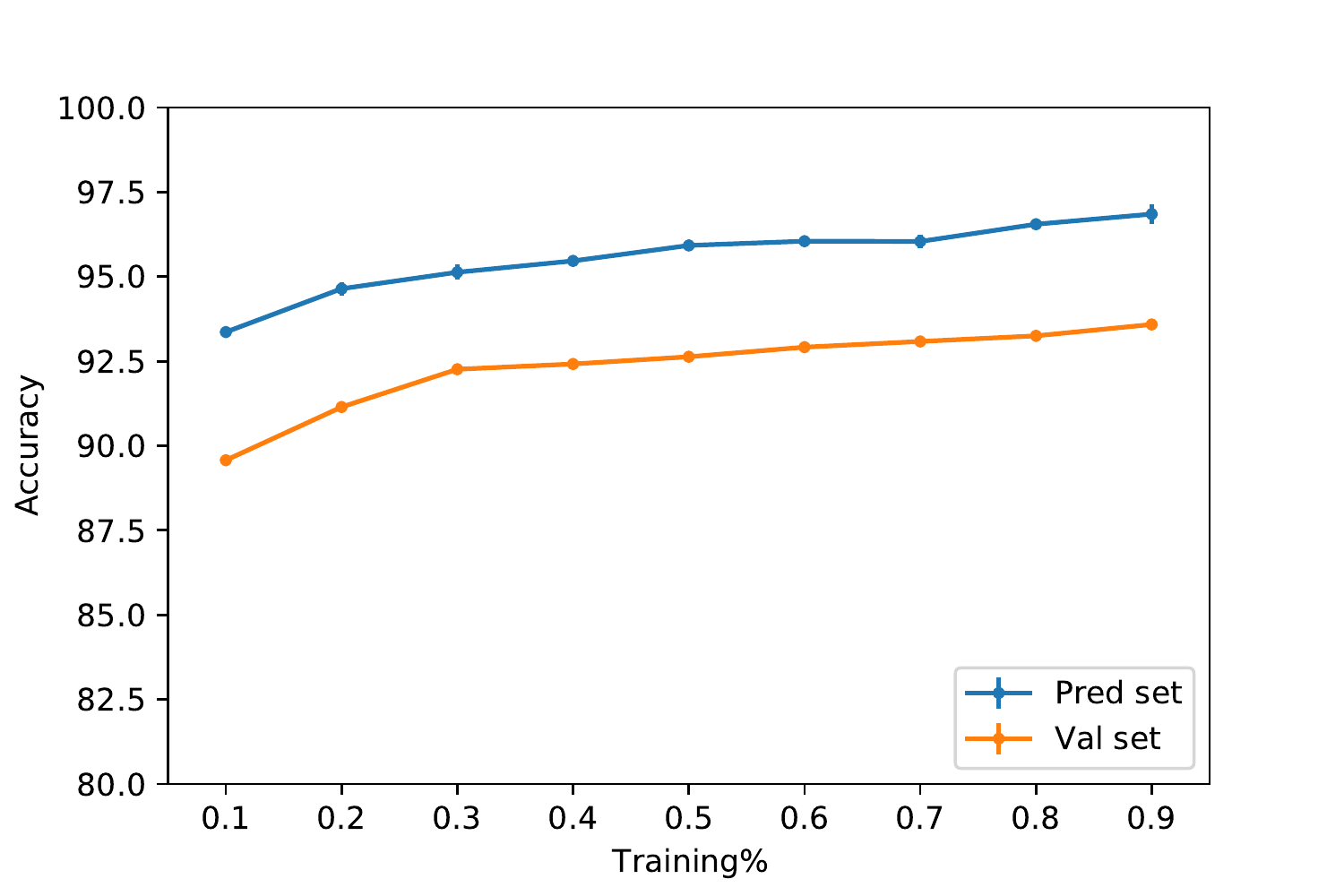}}
\caption{The accuracy of training/testing(left) and prediction(right) evaluated on models trained with different portion of $\cT$. We keep the corresponding validation accuracy on both plots.}
\label{fig:acc/pred22}
\end{figure}

We train our neural network with different portions of $\cT$ and evaluate the trained models on the testing and validation set. 
The training and evaluation process is repeated for $5$ times and the result is plotted in the left of Fig.\ref{fig:acc/pred22}. 
The error bar reflects the stability of our result when we chose different training sets (at a fixed percentage of $\cT$) randomly. 
As before, one can see that the accuracy of the training, testing and validation increases with the size of training data. 
The accuracy on both testing and validation set are above $92\%$. 
This shows that our trained neural network can effectively distinguish the configuration matrices of type $(2,1)$ gCICYs from general $6 \times 3$ matrices. However, the matrices considered here are only for type $(2,1)$ gCICYs with smallest entries $-4$. 
It is more interesting to check if our trained model can predict the gCICYs that are unseen in the data set $\cD$.

\subsubsection*{Prediction}

To do that, we need to construct a new dataset which should contain matrices unseen in the training set $\cD$. 
This new set can be generated in the following steps. 
From the $34,192$ generalized configuration matrices, by taking the parameter to be $5$ or $6$, we can have a much larger set of possible matrices. 
To find gCICYs, we did a scan by computing the line bundle cohomologies of $h^0(\cM,\cL)$ and $h^*(X,\cO)$ for these matrices. 
Unfortunately, time complexity of the computation increases dramatically with the increasing of the entries in the matrix 
\footnote{We impose a constraint in the scan by only considering the cases without involving complicated polynomial computation.}. 
From our preliminary scan, we find $45,259$ gCICYs and $229,540$ non-gCICYs. 
After padding and permutation of rows and the first two columns, we obtain $460,122$ gCICYs and $4,408,940$ non-gCICYs. 
We randomly choose $10,000$ in each case and denoted by $\cP$. 
Notice that, every matrix in the prediction data set $\cP$ contain $-5$ or $-6$ while the smallest entries of matrix in $\cD$ is $-4$. 
Thus, the set $\cP$ is unseen in the training set $\cD$ and can be used for prediction.

We compute the accuracy of our trained model on the prediction set $\cP$ and plot the result on the right of Fig. \ref{fig:acc/pred11}.  
Again, the evaluation is performed on models trained with $5$ different data set in the size.    
We can see that our model can achieve up to $97\%$ accuracy for prediction.
Therefore, we conclude that machine learning is an effective and useful tool to study the classification of gCICY.

With the trained model, we find $16,014$ gCICYs from $92,867$ configuration matrices with most negative entries $-5$ and $20,167$ gCICYs from $125,104$ configuration matrices with most negative entries $-6$. 
Due to the high accuracy for prediction, we believe that our machine learning model can find almost all gCICYs with most negative entries $-5$ and $-6$. 
These predicted gCICYs are attached with this paper.

\section{Conclusion and Outlook}
\label{sec:con}

In this paper, we used machine learning techniques to study the classification of gCICYs. For the data preparation and also for its own usefulness, we fully classified a sub-class of type $(2,1)$ gCICYs, namely, type $(2,1)$ gCICYs with negative entries no smaller than $-4$. This work extends the partial work in this direction initiated in \cite{Anderson:2015iia}.

To show the effectiveness of machine learning, we first used an ANN model with 3 layers to study the fully classified type $(1,1)$ gCICYs. It turns out that, by varying the number of configuration matrices we use and epochs for training, we can get a model which can achieve as high as $96\%$ precision on testing and $98\%$ on validation. Usually in machine learning, the dataset we choose to train the model is just a small portion of all the data. Thus if we get a model which performs very well in the training, this model may not do well for prediction just because it may learn the features belong to the training data only. This phenomenon is called over-fitting in the machine learning literature. However, in our case here, the model we got could achieve $85\%$ precision in prediction. Given the relatively small number of type $(1,1)$ gCICYs and also the simple structure of the neural network we used for this preliminary work, this result already indicates the strong power of machine learning on this kind of classification problem. The high precision on prediction in this part gives us the confidence that machine learning could be a reliable tool in classifying the large number of unknown gCICYs. 

To confirm this idea, we moved on to study the classification of type $(2,1)$ gCICYs. Unlike the case of type $(1,1)$ gCICYs, there are a lot more type $(2,1)$ gCICYs. And as said before, a classification only exists for those with negative entries no smaller than $-4$. So type $(2,1)$ gCICYs is really the ideal arena for us to use machine learning: we have large enough datasets and there is space for machine learning to find something new. In the first part of this study, we trained a neural network model with 6 layers. The data we use are those we classified with entries less than 4. Again in this case, we can get at leat $92\%$ precision on both testing and validation. This result is not that unexpected since given enough data and epochs for training, we could get even higher precision. The problem is, can we generalize this model to something we don't know? And can we really use machine learning technique to predict new knowledge? To answer these two questions, we use the trained model (from the data with negative entries no smaller than $4$) to data with negative entries bigger than $4$. In order to have some sense on how good the model's performance is, we first tried this procedure on some data which we already knew are gCICYs or not and also of course they should have negative entries smaller than $4$. The result of this prediction part is great, $97\%$. This means that the model trained from the data with negative entries no smaller than $4$ should have already learned gCICYs' main common features which can generalize to all gCICYs regardless of the range of the negative entries. Given this fact, we did a prediction by using this model for more gCICYs which are not classified yet. This prediction should have a high precision as indicated by the experiment.

From the experiments we did for type $(1,1)$ and type $(2,1)$ gCICYs, we can see that neural network works very well for the classification of gCICYs. Given that the precision for prediction is so high, especially for type $(2,1)$ gCICYs, we are confident that neural network could actually do more. One possible direction is that neural network may be able to help us to find a quick analytic way to judge if one configuration matrix represents a gCICY or not. Another possible direction is that, we can try to apply other machine learning techniques, like reinforcement learning, to this problem. Unlike the work with neural network, for reinforcement learning the model is not trained by a subset of data first and then generalized to more data. Instead, reinforcement learning will do the work in an un-supervised way and could give us positive examples constantly. So it is natural to see if this technique is more time competitive compared with neural network.

On the other hand, we can use the machine learning technique to explore some substructure of gCICY, especially the discrete symmetry structure considered in both CICY cases \cite{Braun:2010vc,Gray:2021kax} and toric cases \cite{Braun:2017juz}. 
It is interesting to find the $\bZ_2$ orientifold structure \cite{Carta:2020ohw,Altman:2021pyc,Crino:2022zjk} in gCICYs that will be important for type II string compactifications. Besides, the numerical Ricci-flat metric of Calabi-Yau manifolds have been studied for the regular CICY \cite{Anderson:2020hux,Jejjala:2020wcc,Douglas:2020hpv,Larfors:2021pbb,https://doi.org/10.48550/arxiv.math/0512625,Headrick:2005ch, Braun:2007sn,Douglas:2006rr,Douglas:2006hz,Headrick:2009jz,Cui:2019uhy}. We could also study how to extend these methods to compute the numerical metric of gCICYs.

\section*{Acknowledgments}
W. Cui would like to thank Mohsen Karkheiran for the useful discussion.   
J. Wang would like to thank James Gray for the useful discussion at the early stage of this project. X. Gao was supported in part by  NSFC under grant numbers 12005150.
We thank 'DragonTooth' Cluster at Virginia Tech.

\nocite{*}
\bibliography{mlgcicy}
\bibliographystyle{utphys}

\end{document}